\documentclass{article}
\usepackage{amsfonts}

\usepackage[T1]{fontenc}
\usepackage{amsmath}

\input{tcilatex}
\usepackage[T1]{fontenc}
\begin{document}

\date{Nov. 21, 2005}
\title{CR Structures and Asymptotically Flat Space-Times}
\author{Ezra T. Newman$^{1}$ and Pawel Nurowski$^{2}$ \\
$^{1}$Dept of Physics and Astronomy, Univ. of Pittsburgh, Pittsburgh PA. USA
\and $^{2}$Instytut Fizyki Teoretycznej, Uniwersytet Warszawski,ul Hoza 69,
Warszawa, Poland}
\maketitle

\begin{abstract}
We discuss the unique existence, arising by analogy to that in algebraically
special space-times, of a CR structure realized on null infinity for any
asymptotically flat Einstein or Einstein-Maxwell space-time.
\end{abstract}

It has been well-known in a large portion of the relativity community that 
\textit{shear-free null geodesic congruences} play an extremely important
role in general relativity. Though it is hard to argue from a priori
knowledge that this should be true, nevertheless from many examples and
theoretical discoveries, the importance does, in fact, become easy to see.
For example, many of the most important exact solutions of the vacuum
Einstein or Einstein-Maxwell equations possess a degenerate principle null
vector that is both geodesic and shear free, e.g., the Schwarzschild, the
Reissner-Nordstrom, the Kerr and Kerr-Newman metrics. In electrodynamics,
the Lienard-Wiechert \{as well as the complex Lienard-Wiechert\cite
{Sh.Free.Max}\} Maxwell fields have a principle null vector that is also
tangent to a shear free null geodesic congruence. These observations, in
turn, led to the more general issue: find all Einstein metrics that possess
a principle null vector field that is shear-free and geodesic. From this
came the discovery of the algebraically special metrics and the beautiful
Goldberg-Sachs theorem stating that the degenerate principal null vectors 
\textit{for Einstein metrics} are always geodesic and shear free. It opened
the door to the large subject of studying the properties and integrating the
algebraically special metrics. One of the very pretty mathematical
discoveries was the automatic existence of a three-dimensional CR structure%
\cite{LNT,LNT2,LN,Taf,Tr} associated with these metrics. In the special case
of asymptotically flat algebraically special metrics, \{or in their
flat-space limits\} one could choose null infinity, [$\frak{I}^{+}],$ as the
realization space of the CR manifold where a `portion' of the metric (coming
from the congruence itself) defines the CR structure.

From a different perspective Penrose, in his development of flat-space
Twistor Theory, realized the importance of what is now known as the Kerr
theorem, which states that a holomorphic function (of three complex
variables) on projective twistor space, CP$^{3},$ defines in Minkowski
space, a shear-free null geodesic congruence that is, in general, twisting,
i.e., is not surface forming. Thus in some general sense the shear-free null
geodesic congruences lie at the center of Twistor Theory.

\begin{remark}
The real five-dimensional subspace of projective twistor space, $\frak{N}$,
defined by the vanishing of the twistor norm, possesses a five-dimensional
CR structure\cite{PNAT,PC}. The intersection of $\frak{N}$ with the subspace
of CP$^{3}$ that is obtained from the vanishing of a holomorphic function
\{via the Kerr theorem\} is a real three-dimensional CR manifold with the CR
structure inherited from that of $\frak{N.}$
\end{remark}

It is one purpose of this note to point out that the shear-free structures
associated with these special situations, i.e., flat space or the
algebraically special metrics, can be generalized to virtually all
asymptotically flat Einstein or Einstein-Maxwell space-times. Though shear
free null geodesic congruences can not be found in arbitrary space-times,
the idea of \textit{shear-free null geodesic congruences} is generalized to 
\underline{\textit{asymptotically}}\textit{\ shear-free null geodesic
congruences}. They exist in all asymptotically flat space-times.

In several recent articles\cite{NKI,NKII,NKO,NO} we returned, with a rather
unconventional point of view to the study of asymptotically flat solutions
of the Einstein or Einstein-Maxwell equations. The main development in that
work was the realization that for \emph{any} given asymptotically flat
Einstein or Einstein-Maxwell space time with any given Bondi asymptotic
shear (with non-vanishing total electric charge), one could find a class of
asymptotically shear free (but in general twisting) null geodesic
congruences. The class was uniquely given \emph{up to an arbitrary choice of
a complex analytic world-line} in the complex four-dimensional space known
as H-space. In the case of asymptotically flat vacuum space times, by 
\textit{mimicking some terms that are found in algebraically spacial type II
metrics}, this complex world-line can be chosen uniquely. For the case of
the Einstein-Maxwell fields, there are a pair of uniquely defined complex
world-lines in the H-space: one is defined from the Maxwell tensor at null
infinity while the other is found from again mimicking terms found in the
type II metrics. We concentrated on the special or degenerate case where the
two world-lines coincide. At first it was not at all clear as to what
meaning one could assign to this (these) world-line(s). Gradually however
suggestions as to their meaning or physical content did appear. Though it is
not the intent here to go into the details of this issue, we remark that the
real part of the world-line can be identified as a position vector \textit{%
in some sort} of ``screen'' or ''observation'' space, with the meaning of a
``center-of-mass'', while the imaginary part can be identified with the
asymptotically defined specific spin-angular momentum, i.e., spin per unit
mass. The most surprising aspect of this attempt to understand the complex
world-line was discovering its relationship to the Bondi mass-momentum, $%
(M,P^{i})$. Writing the complex world-line as $z^{a}=\xi ^{a}(\tau )=\xi
_{R}^{a}(\tau )+i\xi _{I}^{a}(\tau )$ and with heavy approximations we
discovered that we had the relationship 
\[
P^{i}=M\dot{\xi}_{R}^{i}-\frac{2q^{2}}{3c^{3}}\ddot{\xi}_{R}^{i}-\frac{3}{2c}%
M\epsilon _{ijk}\dot{\xi}_{I}^{j}\dot{\xi}_{R}^{k}+\dots 
\]
Then, using the Bondi mass-momentum loss equation, i.e., the equations for $%
dP^{i}/du,$ the equations of motion for both the real and imaginary parts of
the world-line are determined. One sees immediately that there is now the
classical radiation-reaction force but, in addition, from the mass loss an
additional counter term appears that seems to suppress the runaway solutions
associated with the radiation reaction force.

The main point of this note was however to point out that there is a simple
clear mathematical restatement of our main results\cite{NKI,NKII,NKO,NO}.
What we have shown is that for \emph{any} asymptotically flat
Einstein-Maxwell field (with a non-vanishing charge) there are (in general)
two distinct CR structures given on $\frak{I}^{+}$, one determined from the
Maxwell field, the other from the Weyl tensor. We assumed in that work the
special case where the two world-lines coincided.

The unique CR structure arises in the following manner. We begin with Bondi
coordinates $(u,\zeta ,\bar{\zeta})$ on $\frak{I}^{+}$ and with a Bondi
one-form basis $(n,l,m,\bar{m}).$ The one-form $n$ is the dual to the
tangent vector to the generators of $\frak{I}^{+}$ and $l$ is dual to the
vectors normal to the $u=constant$ slices of $\frak{I}^{+},$ ($m,\bar{m}$
are duals to the tangent vectors of the 'slices', $u=constant$). We then
perform a null rotation around $n$ of the form 
\[
l^{*}=l+\frac{L}{r}\bar{m}+\frac{\bar{L}}{r}m+O(r^{-2}),~~~~m^{*}=m+\frac{L}{%
r}n+O(r^{-2}) 
\]
where $L$ , at this moment, is an arbitrary function on $\frak{I}^{+},$
i.e., $L=L(u,\zeta ,\bar{\zeta}).$ The resulting one-forms on $\frak{I}^{+}$
are (after a conformal rescaling of $m$) 
\begin{eqnarray}
l^{*} &=&du-\frac{L}{1+\zeta \bar{\zeta}}\mathrm{d}\zeta -\frac{\bar{L}}{
1+\zeta \bar{\zeta}}\mathrm{d}\bar{\zeta},~~~  \label{one-forms} \\
~m^{*} &=&\frac{\mathrm{d}\overline{{\zeta }}}{1+\zeta \bar{\zeta}},\qquad ~%
\overline{m}^{*}=\frac{\mathrm{d}{\zeta }}{1+\zeta \bar{\zeta}}.  \nonumber
\end{eqnarray}
We note that, for any choice of the function $L(u,\zeta ,\bar{\zeta}),$ the
three one-forms from Eq.(\ref{one-forms}), are a representative set of
one-forms (up to gauge freedom) that define a CR structure on $~\frak{I}%
^{+}. $

When we require that the new null congruence defined by the $l^{*}$ be
asymptotically shear free, we discover\cite{NKI} that $L=L(u,\zeta ,\bar{%
\zeta})$ satisfies the non-linear differential equation 
\begin{equation}
\text{\dh }L+LL\text{,}_{u}=\sigma (u,\zeta ,\bar{\zeta})  \label{SF}
\end{equation}
with $\sigma $ the freely chosen radiation data; the Bondi asymptotic shear.

This equation can be transformed\cite{NKI}, so that, surprisingly, one could
see immediately that the solutions are given up to the choice of an
arbitrary world-line in H-space. Specifically, the solutions are given in
the following parametric form: Begin with the general solution to the `good
cut equation', $u=X(z^{a},\zeta ,\overline{\zeta })$, with $z^{a}$ an
arbitrary point in H-space. Then chose an arbitrary H-space world-line$,$ $%
z^{a}=\xi ^{a}(\tau )$, and substitute it into solution yielding, $u=X(\xi
^{a}(\tau ),\zeta ,\overline{\zeta })\equiv Z(\tau ,\zeta ,\overline{\zeta }%
).$ Finally by the application of \dh\ to $Z(\tau ,\zeta ,\overline{\zeta }%
), $ we have the solution to Eq.(\ref{SF}) given parametrically by

\begin{eqnarray*}
u &=&Z(\tau ,\zeta ,\overline{\zeta }) \\
L(u,\zeta ,\bar{\zeta}) &=&\text{\dh }Z(\tau ,\zeta ,\overline{\zeta }).
\end{eqnarray*}

We thus have that the family of asymptotically shear-free of null geodesic
congruences are determined by the arbitrary world-line $z^{a}=\xi ^{a}(\tau
) $. At this stage we have a CR structure for every choice of the world-line.

Finally by imposing the conditions (mentioned earlier) on the Weyl tensor
and Maxwell tensor that mimicked the algebraically special Weyl tensor, we
obtained a unique world-line and unique choice of $L(u,\zeta ,\bar{\zeta}).$
Consequently we have a unique CR structure on $\frak{I}^{+}$.

An additional point to note is that if the equation $u=Z(\tau ,\zeta ,%
\overline{\zeta })$ is inverted so that 
\[
\tau =T(u,\zeta ,\bar{\zeta}), 
\]
it becomes easy to show that $\tau $ is a CR function, i.e., that $T(u,\zeta
,\bar{\zeta})$ satisfies the CR equation

\[
\text{\dh }T+LT,_{u}=0 
\]
which gives a geometric meaning to the world-line parameter $\tau .$ The
local \textbf{C}$^{2}$ from which the three-dimensional CR manifold is
defined is ($\tau ,\bar{\zeta}$), with the parametric form of the embedding
given by $(\tau ,\bar{\zeta})=(T(u,\zeta ,\bar{\zeta}),\bar{\zeta}).$

We mention that with our physical interpretation of the world-line, we have
the surprising result that when the two world-lines coincide, i.e., when
there is just one CR structure, one finds that the gyromagnetic ratio, i.e.
the ratio of spin to magnetic moment is the same as that of Dirac, namely $%
g=2$.

A question immediately arises: Is there any analogue of the Kerr theorem,
which gave the shear-free null geodesic congruences in flat-space via
arbitrary holomorphic functions on twistor space, to our case of finding 
\textit{asymptotically} shear-free null geodesic congruences. The answer is
yes. The full details are now being worked out. The role of flat twistor
space is played by asymptotic twistor theory\cite{PM,RP,NPK,Hansen}. The
analogue of the Kerr theorem, which produces asymptotically shear free
congruences, lies in the arbitrary choice of the complex world-line in
H-space. The vanishing of the norm of the asymptotic twistor space\cite{NPK}%
, which produces the analogue of $\frak{N,}$ yields the five-dimensional CR
manifold, $\frak{N}$*, a real subspace of the projective asymptotic twistor
space. The three-dimensional CR structure given on $\frak{I}^{+}$, that has
been described here, is inherited from $\frak{N}$*.

\section{Acknowledgments}

This material is based upon work (partially) supported by the National
Science Foundation under Grant No. PHY-0244513. Any opinions, findings, and
conclusions or recommendations expressed in this material are those of the
authors and do not necessarily reflect the views of the National Science.
E.T.N. thanks the NSF for this support. He also thanks both Roger Penrose
for certain research suggestions and Jerzy Lewandowski for great help in
understanding CR structures. PN was supported by the KBN grant 1 P03B 07529.

\end{document}